\documentclass[showpacs,showkeys,11pt,
preprint,preprintnumbers,nofootinbib,
groupedaddress,superscriptaddress,amsmath,amssymb]{revtex4}

\usepackage[dvipdfm]{graphicx}

\usepackage{cancel,graphics,subfigure}

\usepackage{color}

\def\sla#1{\rlap/#1}

\begin{document}
\title{Search for doubly charged Higgs bosons using the same-sign diboson mode at the LHC}
\author{Cheng-Wei Chiang}
\email{chengwei@ncu.edu.tw}
\affiliation{Department of Physics and Center for Mathematics and Theoretical Physics,
National Central University, Chungli, Taiwan 32001, ROC}
\affiliation{Institute of Physics, Academia Sinica, Taipei, Taiwan 11529, ROC}
\affiliation{Physics Division, National Center for Theoretical Sciences, Hsinchu, Taiwan 30013, ROC}
\author{Takaaki Nomura}
\email{nomura@ncu.edu.tw}
\affiliation{Department of Physics and Center for Mathematics and Theoretical Physics,
National Central University, Chungli, Taiwan 32001, ROC}
\author{Koji Tsumura}
\email{ko2@phys.ntu.edu.tw}
\affiliation{Department of Physics, National Taiwan University, Taipei, Taiwan 10617, ROC} 

\date{\today}

\begin{abstract}
Doubly charged Higgs bosons are predicted in many new physics models with an extended Higgs sector that contains a Higgs triplet field.
Current experimental searches have been focusing mainly on the scenario in which the same-sign dilepton decay modes are the dominant ones. 
We study the scenario where the vacuum expectation value of the triplet field is sufficiently large so that the associated charged Higgs bosons decay dominantly to a pair of weak gauge bosons instead.
A detailed simulation of the signal and the backgrounds is performed for the CERN Large Hadron Collider at the collision energy of $8$ TeV and $14$ TeV.  We find that different cuts should be imposed for the events, depending on whether the doubly charged Higgs boson mass is greater than about 200 GeV.  In the higher mass region, the forward jet tagging proves to be useful in enhancing the signal significance.
We show the discovery reach of the LHC running at 8 and 14 TeV, with two benchmark triplet vacuum expectation values.  With an integrated luminosity of $10$ fb${}^{-1}$ at 8 TeV, the doubly charged Higgs boson with a mass of $\sim 180$ GeV can be tested at $5\sigma$ level in such a scenario. 
\end{abstract}
\maketitle

\section{Introduction} 
The observation of neutrino masses and mixing is well established in the past decade or so~\cite{Ref:solar-v,Ref:atom-v,Ref:acc-v,Ref:shortbaseline-v,Ref:longbaseline-v}.  However, the absolute scale of neutrino masses is still uncertain, and the pattern of neutrino mixing angles is also a mystery in the today's elementary particle physics. 
The smallness of neutrino masses may be related to whether the neutrinos are Dirac or Majorana fermions.  The fact that the neutrino mixing can be approximated by the tri-bi-maximal mixing~\cite{Ref:TBM} may suggest some underlying structure such as flavor symmetry~\cite{Ref:Fsym}. 

In the standard model (SM), neutrinos are assumed to be massless due to the absence of right-handed neutrinos. 
In order to explain the tiny neutrino masses, many mass generation mechanisms have been proposed in the literature~\cite{Ref:Type-I,Ref:Type-II,Ref:Type-III,Ref:Zee,Ref:Zee-Babu,Ref:Ma}. 
Upon introducing right-handed neutrinos, one can have Dirac mass terms for neutrinos. 
However, neutrinos have unnaturally small masses as compared to the other SM fermions. 
Therefore, one may expect that neutrinos are in fact Majorana particles and that their masses are generated from a different origin~\cite{Ref:Type-I,Ref:Type-II,Ref:Type-III,Ref:Zee,Ref:Zee-Babu,Ref:Ma} than the other SM fermions. 
There are three possibilities to generate the Majorana mass for neutrinos if one is allowed to introduce only one type of new particles at a time.  They are generally dubbed Type-I, -II, and -III seesaw mechanisms in which the SM is added with $SU(2)_L$ singlet fermions~\cite{Ref:Type-I}, triplet scalar bosons~\cite{Ref:Type-II}, and triplet fermions~\cite{Ref:Type-III}, respectively.  The Majorana mass term thus generated involves a dimension-5 operator and is thus suppressed by some dimensionful parameter in the model. 

The Type-II seesaw mechanism is often called the Higgs triplet model (HTM), 
where the neutrino masses are generated by the vacuum expectation value (VEV) of 
the neutral component of a complex Higgs triplet field~\cite{Ref:Type-II}. 
An important feature of the HTM is that the neutrino mass matrix is directly 
related to the Yukawa coupling matrix of the Higgs triplet field.  Another characteristic feature of the model is the existence of a doubly charged Higgs boson.  In a good portion of the triplet VEV range, the particle decays dominantly into a pair of same-sign leptons, leading to distinctive signatures at hadron colliders.  Moreover, such leptonic decays may involve flavor violation and have different branching ratio patterns, depending on the triplet VEV, Majorana phases, and the neutrino mass spectrum \cite{Ref:AC}.  Thus, a detailed study of the leptonic decays of the doubly charged Higgs boson may provide useful information on the neutrino mass matrix~\cite{Ref:mass_matrix}, giving us a possibility to reveal the nature of neutrinos through collider studies~\cite{Ref:ACG}.

The doubly charged Higgs boson has been searched for at the Tevatron~\cite{Ref:H++TeV} and the CERN Large Hadron Collider (LHC)~\cite{Ref:H++LHC}.  In the experimental analyses, it is generally assumed that the doubly charged Higgs boson dominantly decays into same-sign lepton pairs with the same or different flavors.  Lower mass bounds are then set by further assuming specific leptonic decay patterns.  Besides such a scenario, we find no other mass bound available from hadron colliders otherwise.  The precision measurements of the $Z$ boson decay width at the LEP have constrained the doubly charged Higgs boson mass to be above $m_Z^{}/2$~\cite{Ref:H++LEP}. \footnote{
Even at $e^+e^-$ colliders, the doubly charged Higgs boson with a mass more than $m_Z^{}/2$ can be probed by looking for the same-sign dilepton signal via 
$H^{++} H^{--} \to 4j + 2\ell + \cancel{E}$ where one of the doubly charged Higgs boson may be off mass shell.}

In this paper, we study the scenario where the triplet VEV is sufficiently large so that the doubly charged Higgs bosons decay dominantly into a pair of weak gauge bosons instead.
Even though one still looks for same-sign leptons from the $W$ boson decay, this case is more involved because of the missing energy in the final state and suffers from larger SM backgrounds.  We perform a detailed simulation for both the HTM and the model proposed by Georgi and Machacek~\cite{Ref:GM}. 
We here consider the range of the triplet VEV between $1$ and a few tens of GeV which is allowed by experimetal data, and  
analyze the phenomenology of such a doubly charged Higgs boson with a mass of ${\cal O}(100~{\rm GeV})$ and examine the LHC reach by imposing appropriate cuts.

The structure of this paper is organized as follows.  In Sec.~\ref{sec:model}, we brief review  the HTM and Georgi-Machacek model as well as introduce the framework for later analyses.  In Sec.~\ref{sec:DCH}, we present the properties of the doubly charged Higgs boson in the large triplet VEV scenario, including its total decay width and cross sections of various channels at the LHC.  Sec.~\ref{sec:simulation} gives details of collider simulations of both signal and background events.  We divide our analysis of no forward jet tagging into high-mass and low-mass regions and discuss the corresponding cuts.  In the high-mass region, we further propose the use of forward jet tagging to improve the significance of the signal.  We show various distributions for the collision energy of 8 TeV only, noting that those for 14 TeV are qualitatively the same but differ in magnitude.  Finally, we show the required luminosity for a 5-sigma discovery of the doubly charged Higgs boson for two benchmark triplet VEV's.  Our findings are summarized in Sec.~\ref{sec:summary}.

\section{Models with doubly charged Higgs bosons
\label{sec:model}}

Models with a doubly charged Higgs boson have been introduced some time ago~\cite{Ref:Type-II,Ref:Zee-Babu,Ref:LH,Ref:LR,Ref:Composite}.  The HTM with the Type-II seesaw mechanism introduces a complex Higgs triplet field which contains the doubly charged Higgs boson~\cite{Ref:Type-II}.  One important function of such a field is to give tiny Majorana-type masses to neutrinos through the VEV of its neutral component.
A two-loop radiative seesaw model for neutrinos proposed by Zee and Babu also involves 
an $SU(2)$-singlet doubly charged Higgs boson~\cite{Ref:Zee-Babu}.
To stabilize the Higgs boson mass, the littlest Higgs model also invokes doubly charged Higgs bosons in the enlarged Higgs sector~\cite{Ref:LH}. 
In the left-right symmetric model, an $SU(2)_L$ triplet scalar is introduced 
as a partner of the $SU(2)_R$ triplet that breaks the left-right symmetry~\cite{Ref:LR}. 
The doubly charged Higgs boson is also predicted in certain composite Higgs models~\cite{Ref:Composite}.
Recently, the authors of Ref.~\cite{Rentala:2011mr} have performed a systematic study of the leptonic channels of a doubly charged Higgs boson in the $SU(2)_L$ singlet, doublet, and triplet representations at the LHC.

In the HTM, an $SU(2)_L$ complex triplet scalar $\Delta$ with $Y=1$ is introduced to the SM
\begin{align}
\Delta = \begin{pmatrix}
\Delta^+/\sqrt2 & \Delta^{++} \\ \Delta^0 & -\Delta^+/\sqrt2
\end{pmatrix},
\end{align}
where $\Delta^{++}$ is the doubly charged Higgs boson. 
The triplet VEV, $v_\Delta^{} (\equiv \sqrt2 \langle \Delta^0 \rangle)$, breaks 
the custodial $SU(2)$ symmetry at tree level, and gives the deviation of 
the electroweak rho parameter from unity as
$\rho = \frac{1+2x^2}{1+4x^2}$~\cite{Ref:rho},
where $x=v_\Delta^{}/v$ where $v$ is the VEV of the $SU(2)$ doublet Higgs field 
which satisfies $\sqrt{v^2+2v_\Delta^2} \simeq 246$ GeV. 
The triplet VEV is constrained to be less than $4$ GeV at the $95 \%$ confidence level \cite{Ref:PDG} 
\footnote{
In a recent study~\cite{Ref:EWHTM}, a one-loop level analysis of the electroweak precision data is performed for the HTM, taking 
$\alpha_{\text{em}}$, $G_F$, $m_Z$ and $\sin^2\theta_W$ as the input patameters.
They find that the data allow a triplet VEV up to ${\cal O}(10 {\rm GeV})$, depending on the mass splitting between the charged Higgs bosons and the doubly charged Higgs boson mass is preferably to be less than $200$ GeV.
}. 
The Majorana mass of neutrinos can then be generated $(M_\nu)_{\ell\ell'} = \sqrt2 h^{}_{\ell\ell'} v_\Delta$ 
by the Yukawa interaction of the triplet field as 
\begin{align}
{\mathcal L}_{\rm Yukawa} &= h_{\ell\ell'}^{} 
\overline{L_\ell^c} \, \Delta\, i\, \sigma_2 L_{\ell'} +\text{h.c.}.
\end{align}
Since the mass scale of neutrinos is expected to be less than $1$ eV, 
the Yukawa coupling constants $h^{}_{\ell\ell'}$ can be of order one if 
$v_\Delta^{} \sim 1$ eV. 
These Yukawa coupling constants can induce lepton flavor-violating processes. 
Hence, the combination of the doubly charged Higgs boson mass and the triplet VEV 
(roughly $\sim m_{\Delta^{\pm\pm}} v_\Delta$) can be constrained by LFV data~\cite{Ref:HTM-LFV}. 
In order to satisfy the LFV constraint, either heavy triplets or large triplet VEV 
(or both) is required. In our present study, we will take relatively large triplet 
VEV and therefore LFV bounds are negligible.
The weak interaction of the doubly charged Higgs boson in the HTM is given by
\begin{align}
{\mathcal L}_\text{weak}^\text{HTM} &\simeq 
+g^2 \frac{v_\Delta^{}}{\sqrt{2}} \Delta^{++}W_\mu W^\mu 
+g W^\mu (\Delta^- \partial_\mu \Delta^{++} -\Delta^{++} \partial_\mu \Delta^-)
+ \text{h.c.},
\label{Eq:HWW}
\end{align}
where ${\mathcal O}(v_\Delta/v)$ terms are neglected.

In the Georgi-Machacek (GM) model~\cite{Ref:GM}, the additional real triplet field 
$\chi = (\chi^-,\chi^0,\chi^+)$ is required to have the aligned VEV 
as the complex triplet field, keeping the $\rho$ parameter unity at tree level. 
The triplet VEV in this model, however, is constrained to be $\lesssim 55$ GeV for 
$m_{H^{\pm\pm}}^{} \sim 100$ GeV by the $Z\to b\bar b$ data~\cite{Ref:GM-VEV}.  
We note in passing that in fact the upper bound on the triplet VEV has some dependence 
on the doubly charged Higgs boson mass, as shown in Ref.~\cite{Ref:GM-VEV}.  
It can be bigger for larger mass values. 
The Yukawa interaction of the doubly charged Higgs boson is same as in the HTM,
and the weak interaction with the singly charged triplet Higgs boson is shared by 
two mass eigenstates as
\begin{align}
{\mathcal L}_\text{weak}^\text{GM} &\simeq 
+g^2 \frac{v_\Delta^{}}{\sqrt{2}} \Delta^{++}W_\mu W^\mu 
+\frac1{\sqrt2} g W^\mu \bigl[
(H_1^--H_2^-) \partial_\mu \Delta^{++} -\Delta^{++} \partial_\mu (H_1^--H_2^-)\bigr]
+ \text{h.c.},
\label{Eq:HWW-GM}
\end{align}
where $H_{1,2}^\pm \simeq (\chi^\pm \pm \Delta^\pm)/\sqrt2$.

For $v_\Delta^{} \lesssim 10^{-4}$ GeV, the doubly charged Higgs boson mainly 
decays into same-sign lepton pairs~\cite{Ref:Type-II-LHC}. 
On the other hand, for relatively large values of the triplet VEV ({\it e.g.},
$v_\Delta \simeq 5$ GeV), the Yukawa coupling constants must be smaller 
than $10^{-9}$, and the doubly charged Higgs boson predominantly decays
into a same-sign $W$ boson pair \cite{Ref:Type-II-LHC, Ref:Roadmap}. 

The doubly charged Higgs boson has been directly searched for at the LHC 
by looking for the pair production process $pp\to \Delta^{++}\Delta^{--}$ 
and also the associated production with the singly charged Higgs boson
$pp\to \Delta^{\pm\pm}\Delta^{\mp}$~\cite{Ref:AC,Ref:H++H--}. 
In these analyses, decays of the charged Higgs bosons are assumed to be purely leptonic, and hence the lower mass bound thus obtained depends a lot on the leptonic decay pattern~\cite{Ref:H++TeV,Ref:H++LHC}.  At any rate, it is obvious that these results are not applicable for $v_\Delta^{} \agt 10^{-4}$ GeV. 
In order to constrain the doubly charged Higgs boson in this case, a different strategy is required at hadron colliders.

\section{Doubly charged Higgs bosons at the LHC
\label{sec:DCH}}

\begin{figure}[t]
\centering
\includegraphics[scale=0.85]{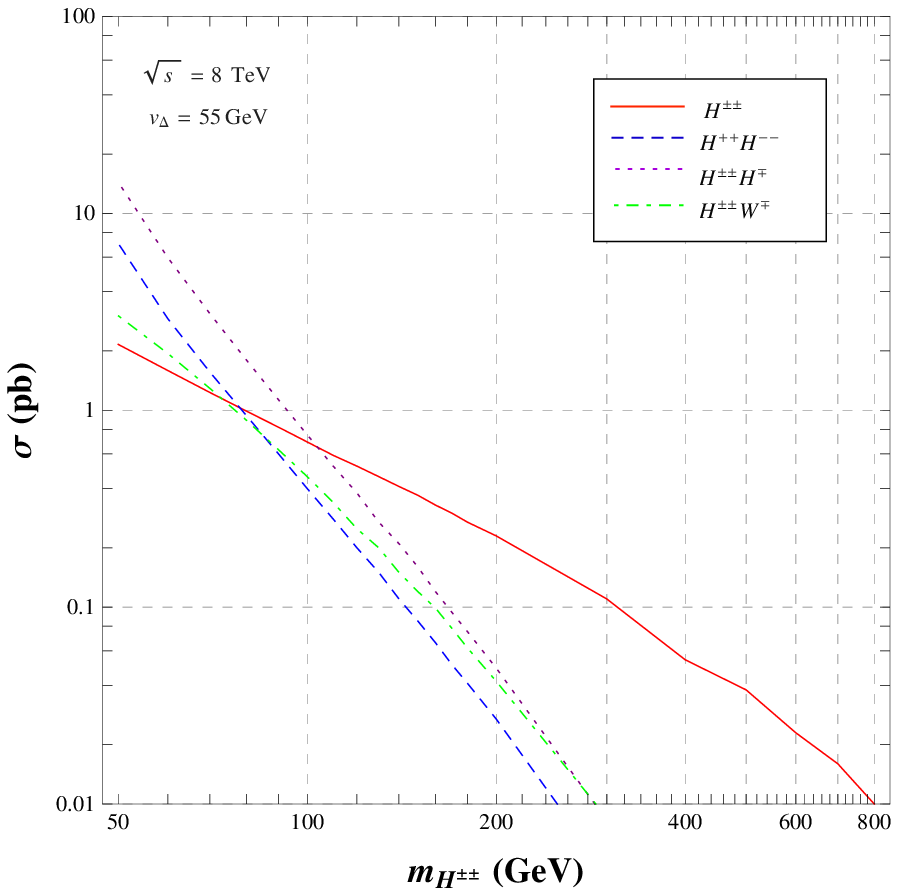}
\includegraphics[scale=0.85]{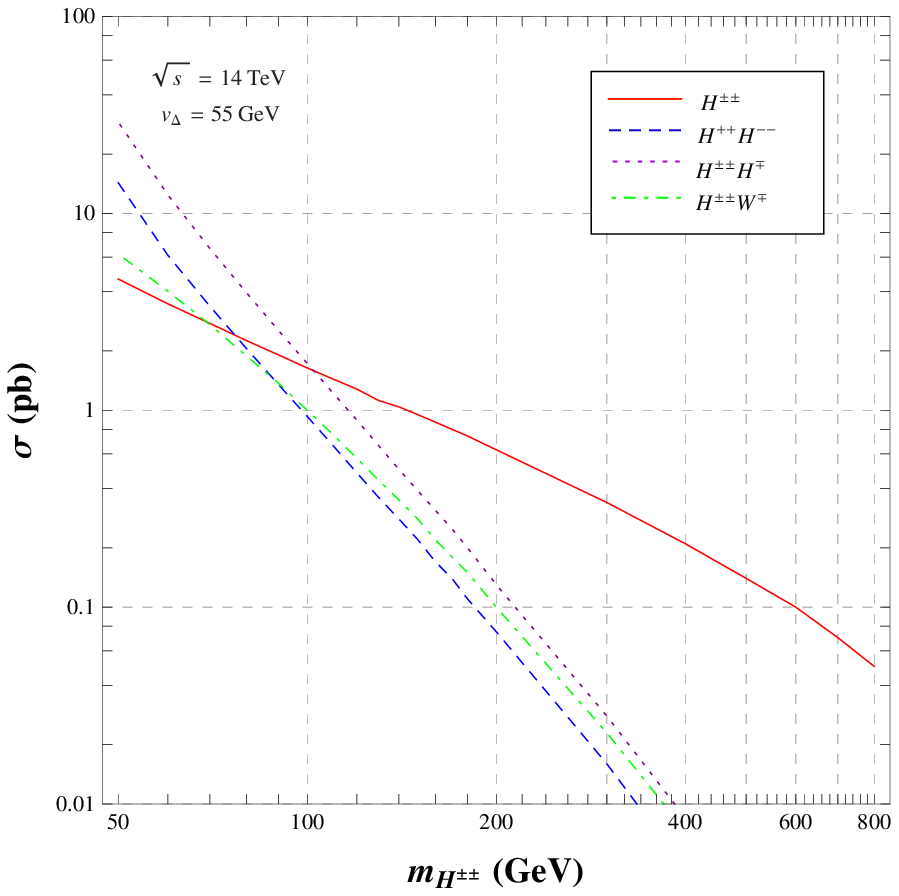}
\caption{Cross sections of the doubly charged Higgs bosons in various production channels for $v_{\Delta} = 55$ GeV. 
The collision energy is assumed to be $8$ TeV in the left plot and $14$ TeV in the right plot.  The {\tt CTEQ6L} PDF's are used.
\label{FIG:Xsec}}
\end{figure}
%

In this section, we study the production channels and the decays of the doubly 
charged Higgs boson for the scenario where the scalar triplet field acquires a relatively large VEV, $v_\Delta^{}$. 
With $v_\Delta \sim {\cal O}{\rm (GeV)}$ to a few tens of GeV, 
the weak interaction of the doubly charged Higgs boson in Eq.~\eqref{Eq:HWW} 
can induce significant effects in hadron collider phenomenology.  Hereafter, we denote generic doubly and singly charged Higgs bosons by $H^{\pm \pm}$ and $H^\pm$, respectively,
in order to distinguish them from $\Delta^{\pm \pm}$ and $\Delta^{\pm}$ defined in the previous section for the HTM.  However, we still use $v_\Delta$ to denote the triplet VEV.  For simplicity and definiteness, we will assume $H^{\pm\pm}$ and $H^{\pm}$ are degenerate in our simulations.

In FIG.~\ref{FIG:Xsec}, we show the cross sections of various production channels for 
the doubly charged Higgs bosons in the HTM at the LHC as a function of $m_{H^{\pm\pm}}^{}$, 
the mass of $H^{\pm\pm}$.  The collision energy is set at $8$ TeV and $14$ TeV. 
The {\tt CTEQ6L} parton distribution functions (PDF's) \cite{Ref:CTEQ} are used. 
For a relatively light doubly charged Higgs boson, scalar pair production processes 
such as $pp\to Z/\gamma \to H^{++}H^{--}$ and $pp\to W \to H^{\pm\pm}H^{\mp}$ give the 
largest cross sections. 
Note that these production channels are independent of the triplet VEV. 
In the GM model, there are two singly charged Higgs bosons which interact with 
the doubly charged Higgs boson with suppressed coupling by factor $1/\sqrt2$. 
Assuming mass degeneracy between two the singly charged scalars, 
the sum of the $H^{++} H_1^-$ and $H^{++} H_2^-$ production 
coincides with the cross section in the HTM. 
For a relatively heavy doubly charged Higgs boson, 
the single production via the same-sign $W$ boson fusion process can be a 
dominant channel when the Higgs triplet VEV is taken to be, for example, $55$ GeV, the upper limit in the GM model with $m_{H^{\pm\pm}} \sim 100$ GeV.  The explicit value of $m_{H^{\pm\pm}}$ above which the $W$ boson fusion process becomes dominant will shift higher (lower) for a smaller (larger) $v_\Delta$.  As we will see later, this affects our choice of whether to impose the forward jet tagging cut.
The cross sections are simply proportional to $v_\Delta^2$, 
and decreases more slowly with $m_{H^{\pm\pm}}$ than the other channels. 
The doubly charged Higgs boson can also be radiated off the $W$ boson: 
$pp\to W^{\pm} \to H^{\pm\pm}W^{\mp}$.  These production cross sections are also proportional to $v_\Delta^2$. 
In calculating the cross sections, we have applied the gauge interactions of the triplet Higgs bosons in Eq.~\eqref{Eq:HWW} for the HTM besides the degenerate spectrum of the triplet Higgs bosons. 
%
%
We also note that if there is a mass hierarchy among the triplet Higgs bosons: $m_{H^{\pm \pm}} < m_{H^{\pm}} < m_{H^{0}}$, 
the $H^{\pm \pm}$ production cross section can be significantly increased \cite{Ref:Roadmap,Chun:2003ej,Ref:MassHierarchyCase1}.  
Our work serves as a more conservative study.

\begin{figure}[t]
\includegraphics[scale=0.85]{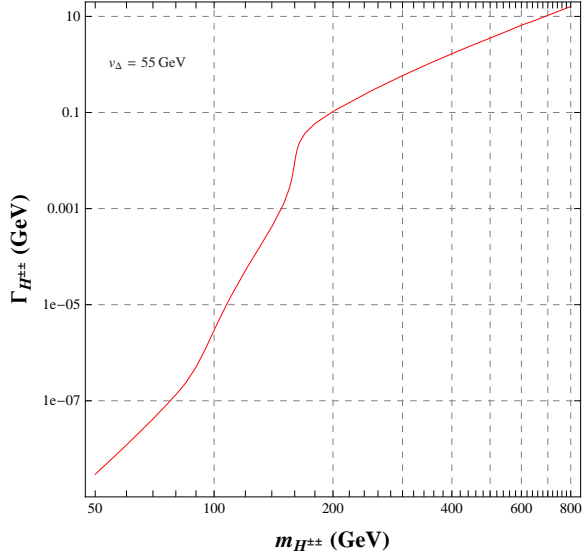}
\caption{Partial decay width of $H^{\pm \pm} \to W^\pm W^\pm$ as a function of $m_{H^{\pm\pm}}$.  The computation assumes $v_\Delta^{} = 55$ GeV, and is done with CalcHEP 3.1 \cite{Ref:CalcHEP}.}
\label{width}
\end{figure}
%

Once the doubly charged Higgs bosons are produced, they decay mainly into a pair of same-sign $W$ bosons so long as the triplet VEV is much larger than $10^{-4}$ GeV. 
In FIG.~\ref{width}, we show the partial decay width of the doubly charged Higgs boson into a $W$ boson pair as a function of its mass.  For definiteness, $v_\Delta^{}=55$ GeV is taken.  For a different value of $v_\Delta$, the width can be simply scaled with $v_\Delta^2$. 
For $v_\Delta^{} \gg 10^{-4}$ GeV, this partial width is a good approximation to 
the total width of the doubly charged Higgs boson in the HTM.  
For $m_{H^{\pm\pm}}$ below the $WW$ threshold, one or both $W$ bosons must be off shell.  However, $WW$ is still the main decay mode because the magnitude of the 
leptonic Yukawa interactions is determined by $m_\nu/v_\Delta^{}$ to be small.
Note that if one has the mass hierarchy $m_{H^{\pm \pm}} > m_{H^{\pm}} > m_{H^{0}}$, the $H^{\pm \pm} \to H^{\pm}W^{\pm}$ decay mode is also relevant \cite{Ref:Roadmap,Ref:MassHierarchyCase2}.
A study of the $H^{\pm\pm}$ decay pattern under such a mass hierarchy is given recently in FIG.~1 of \cite{Ref:Roadmap}.

\section{Simulation studies
\label{sec:simulation}}

\subsection{Framework of event generation and pre-selection}

We here focus on the signals from various doubly charged boson production mechanisms
with a relatively large triplet VEV, including
the scalar pair production, $pp\to H^{++}H^{--},H^{\pm\pm}H^{\mp}$, 
the vector boson fusion (VBF) process, $pp\to H^{\pm\pm}jj$ \cite{Ref:H++VBF}, and 
the associated production, $pp\to H^{\pm\pm}W^{\mp}$.  Only one of the doubly charged Higgs bosons in these processes decays to a pair of $W$ bosons, each of which then decays leptonically.
A pair of same-sign dileptons with relatively large $p_T^{}$ are required in the final state in order to identify the lepton number violating signal.  Hereafter, the {\it lepton}, denoted by $\ell$, refers to the electron or muon.  Our calculation of
$H^{\pm \pm} \rightarrow \ell^\pm \ell^\pm \sla{E}$ includes the chain decay $W \rightarrow \tau \nu \rightarrow \ell \nu \nu \nu$.
%
For the VBF signal in particular, the events contain large missing $p_T^{}$ and two high-$p_T^{}$ jets in the forward region. 
The signal events are generated using {\tt MadGraph/MadEvent}~\cite{Ref:MG}, 
and then passed to {\tt PYTHIA} \cite{Ref:Pythia} in order to include initial-state radiation (ISR) and final-state radiation (FSR) effect. 
We use the {\tt CTEQ6L} PDF's and assume the collision energy of 8 TeV and 14 TeV. 
The detector level simulations are carried out with the {\tt PGS} package \cite{Ref:PGS}. 
Two benchmark values, $1$ GeV and $55$ GeV, are used for the triplet VEV.

In our simulation studies, we take into account the following background processes:
\begin{itemize}
\item Drell-Yan (DY) background:
$p p \to Z^{(*)}(\to \ell^+ \ell^-) +\text{ISR/FSR leptons} \,  (\ell=e, \mu)$;
\item Electroweak (EW) background of ${\mathcal O}(\alpha^4)$: 
$p p \to W^{\pm} W^{\pm} j j$;
\item QCD background of ${\mathcal O}(\alpha^2 \alpha_s^2)$:
$p p \to W^{\pm} W^{\pm} j j$;
\item top quark background:
$p p \to W^{\pm} t \bar{t}, W^{\pm} t \bar{t} + j$; and
\item $VV (V=W$ or $Z)$ background: 
$p p \to VV + nj \quad (n \leq 2)$.
\end{itemize}

The DY background consists of the leptonic $Z$ decay with ISR/FSR leptons.  The highest $p_T^{}$ lepton in such events may be mis-identified as the signal 
coming from the doubly charged Higgs boson with a similar mass. 
However, we expect that the requirement of a sufficiently large $p_T^{}$ 
for the second highest $p_T^{}$ lepton of the same sign should be effective in reducing 
the background.  This is because the second highest $p_T^{}$ same-sign lepton 
in this background originates from ISR/FSR leptons and thus tends to be soft. 

The EW background is generated by the $t$-channel $Z/\gamma$ exchange 
in same-sign $WW$ scattering. 
This background process can interfere with the signal process of the $s$-channel VBF. 
However, it can be reduced by constructing a kinematical variable such as 
the invariant mass or the cluster transverse mass, as the on-shell 
production of doubly charged Higgs bosons will dominate.

The QCD background involves color exchange between the initial 
and final partons, with the weak bosons being emitted from these partons. 
Because of the large QCD activities, the outgoing partons are no longer 
isolated and the background can be rejected by forward jet tagging.

The top quark background has top quarks in the final state along 
with a $W$ boson and possibly more jets. 
This process can produce a pair of the same-sign $W$ bosons, one of which comes from the top quark decay. 
This background can also be reduced by the requirement of the high $p_T^{}$ leptons. 
For the VBF signal, these background events can be 
further reduced by forward jet tagging.

The $VV$ background from $ZW$ and $ZZ$ production and with one or two jets may produce relatively hard same-sign dilepton events. 
The events can also be reduced by forward jet tagging.

The signal of $H^{\pm\pm}$ cannot be fully reconstructed due to the neutrinos in the final states.  In the case where $H^{\pm\pm}$ is heavy and the VBF process dominates, however,
it is useful to consider the cluster transverse mass variable~\cite{Ref:CTM} of the $WW$ pair, constructed from the dilepton momentum and the missing transverse momentum.
Explicitly, the cluster transverse mass is defined as 
\begin{equation}
M^2_c(WW) = \Bigl[ \sqrt{M^2_{\ell \ell}+({\vec p}_T^{\, \ell\ell})^2} + |\cancel{\vec p}_T|^2 \Bigr]^2 
-\Bigl[ {\vec p}_T^{\, \ell \ell} +\cancel{\vec p}_T^{} \Bigr]^2,
\label{Eq:MC}
\end{equation} 
where $M_{\ell \ell}$ is the invariant mass of the two charged leptons, 
${\vec p}_T^{\, \ell \ell}$ is the vector sum of their transverse momenta, and 
$\cancel{\vec p}_T^{}$ is the missing transverse momentum determined 
by the negative sum of visible momenta in the transverse direction.  
The endpoint in the distribution of $M_c(WW)$ corresponds to the mass of the doubly charged Higgs boson.

Throughout this paper, the signal significance is evaluated using \cite{Ref:significance}
\begin{equation}
{\mathcal S} = \sqrt{2 [(s+b) \ln (1+s/b)-s ]},
\end{equation}
where $s$ and $b$ denote the numbers of signal and background events, respectively. 
In the limit of $s/b \ll 1$, it reduces to the simple estimator $s/{\sqrt{b}}$

We first perform the pre-selection of events with the same-sign dileptons using {\tt PGS}. 
In order to simulate the detector performance, leptons are required to be 
isolated and have $p_T^{}>15$ GeV and $\eta<2.5$, where $\eta = \tfrac12\ln\tan\tfrac{\theta}2$ is the pseudorapidity defined using the scattering 
angle $\theta$ in the laboratory frame. 
Jets are constructed using the cone algorithm with $R=0.5$ and are required to have 
$p_T^{}>20$~GeV and $\eta < 5.0$.

\subsection{Event analysis without forward jet tagging}

If the doubly charged Higgs boson is light, they will be dominantly produced in scalar boson pairs such as $H^{++}H^{--}$ 
and $H^{\pm\pm}H^{\mp}$ at the LHC.  Assuming mass degeneracy between $H^{\pm\pm}$ and $H^\pm$, the latter cross section is slightly larger.
Even when the mass of the doubly charged Higgs boson is heavy, 
the pair production process would be dominant for $v_\Delta$ below a few GeV. 
In order to extract the signal events from the backgrounds, 
we require same-sign dileptons with high transverse momenta.  We first do not impose forward jet tagging in the event selection.
The mass of the doubly charged Higgs boson is taken to be 
$90$, $150$ and $200$ GeV for the low-mass analysis and 
$200$, $300$ and $500$ GeV for the high-mass analysis. 
The triplet VEV $v_\Delta^{}$ to be $55$ GeV and the $pp$ collision energy 
of $8$ TeV is assumed in this subsection\footnote{The upper bound on the triplet VEV in the GM model can be weakened for larger masses of additional scalar bosons.}. 
A wider parameter space will be discussed later. 

For the pair production signal, we expect that one doubly charged Higgs boson 
decays into the same-sign dileptons with neutrinos and the other doubly or 
singly charged Higgs decays hadronically through weak gauge bosons. 
Therefore, we require two same-sign leptons and two or more jets with 
no other high $p_T^{}$ leptons as our pre-selection cut. 

%
\begin{figure}[tb]
\centering
\includegraphics[scale=.85]{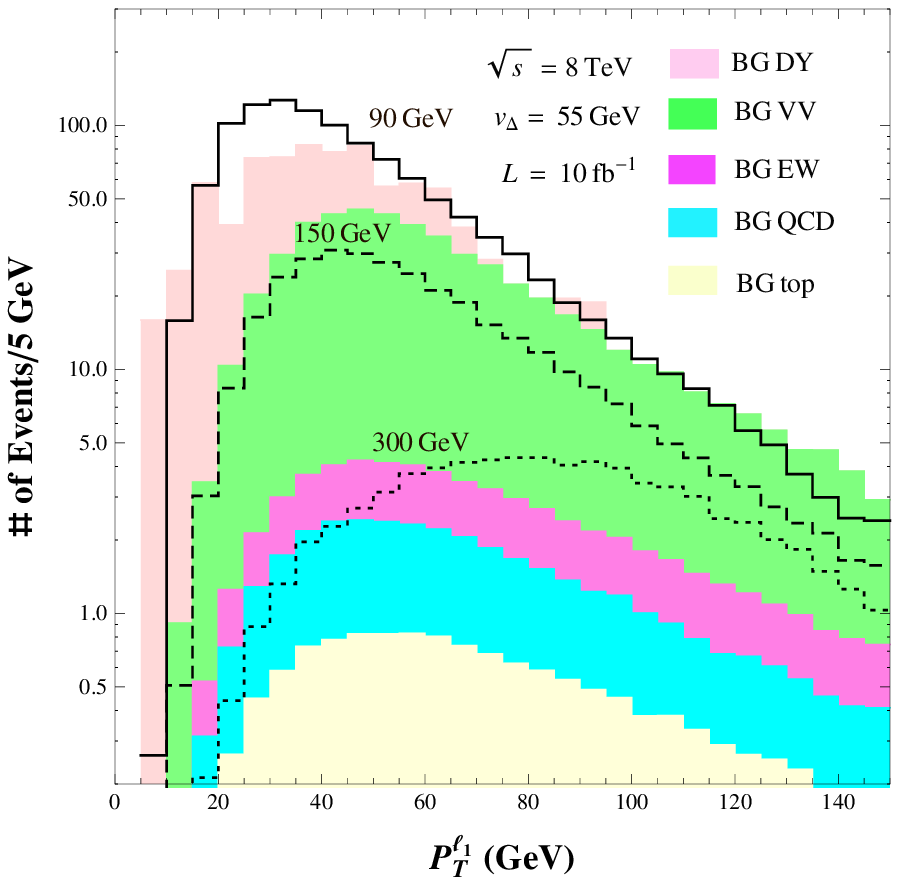}
\hspace{0.5cm}
\includegraphics[scale=.85]{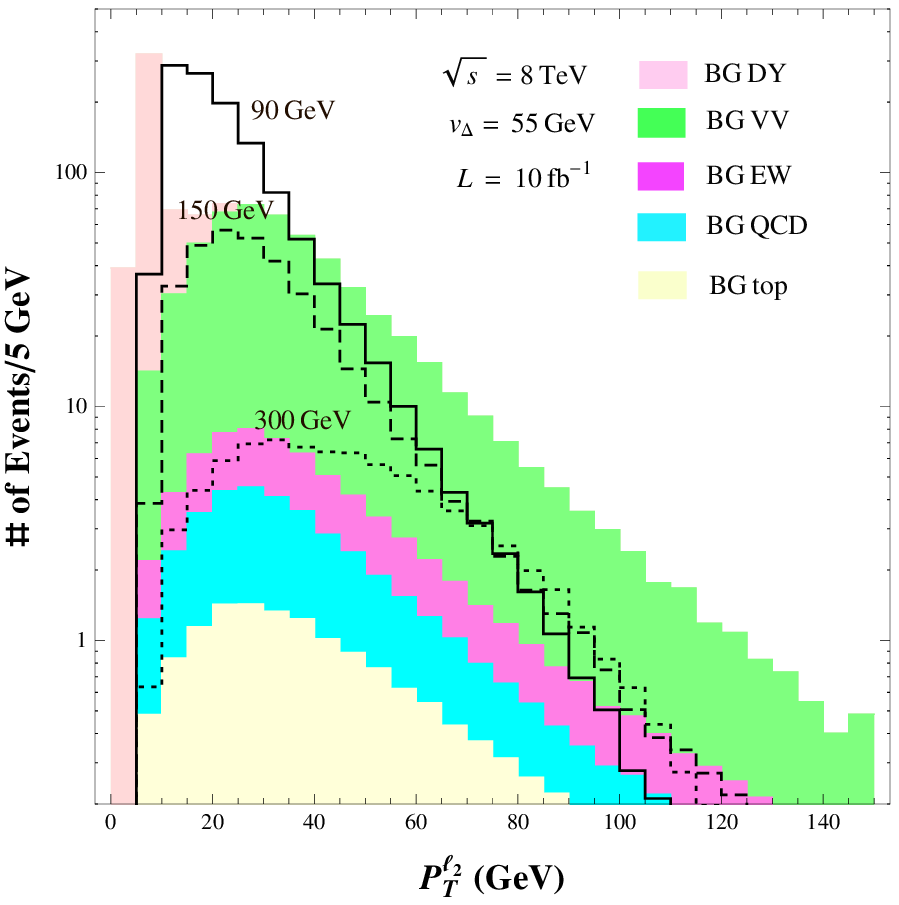}
\caption{The transverse momentum distributions of the highest and the second highest 
leptons of the same-sign dileptons in the signal events with 
$m_{H^{\pm\pm}}=90, 150, 300$ GeV and various SM backgrounds, 
where we choose $v_\Delta=55$ GeV and the LHC collision energy to be $8$ TeV. }
\label{FIG:pTdist}
\end{figure}

%
In the following, $\ell_{1,2}$ denote the same-sign leptons with the highest and second highest $p_T^{}$, respectively.  The DY process with ISR/FSR is the biggest background with the same-sign dileptons.  Since the ISR/FSR leptons are expected to be soft, $p_T^{\ell_2}$ are mostly small in such events.
In FIG.~\ref{FIG:pTdist}, we show the $p_T^{}$ distributions for 
$\ell_1$ (left panel) and $\ell_2$ (right panel) for signal and background events before applying the $p_T^{}$ cuts. 
It is clearly that the signal events can be readily discriminated from the DY background 
by the $p_T^{}$ cuts for the outgoing leptons.

\subsubsection{Low-mass analysis}

\begin{figure}[t]
\centering
\includegraphics[scale=.85]{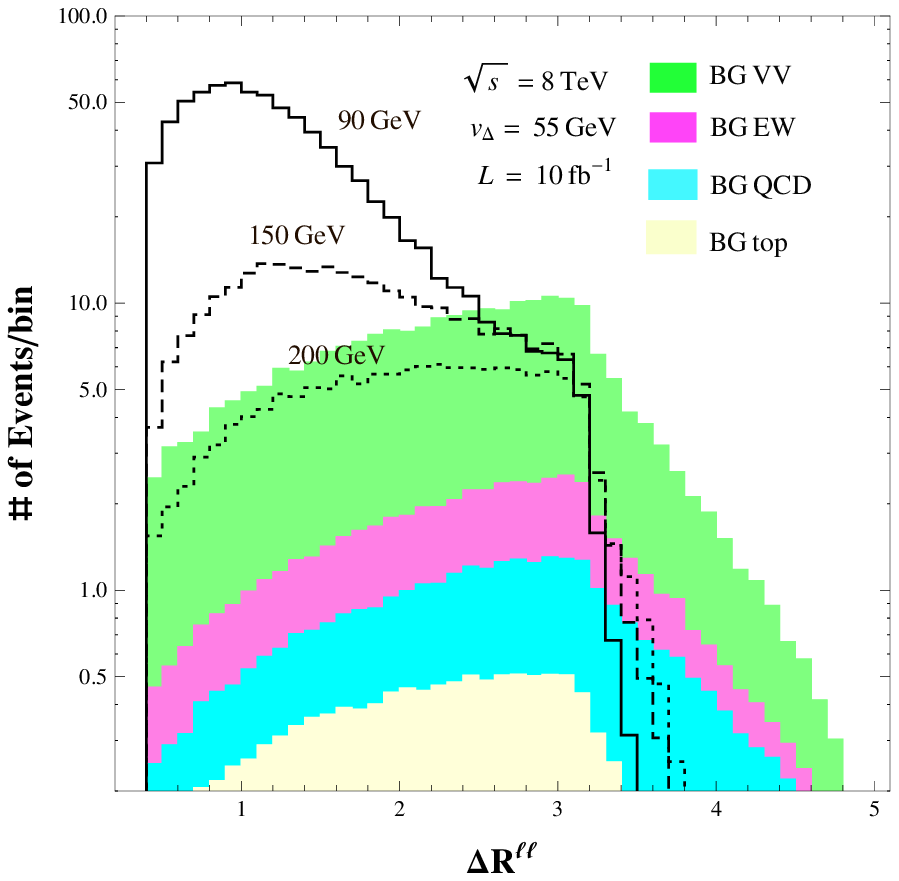}
\hspace{0.5cm}
\includegraphics[scale=.85]{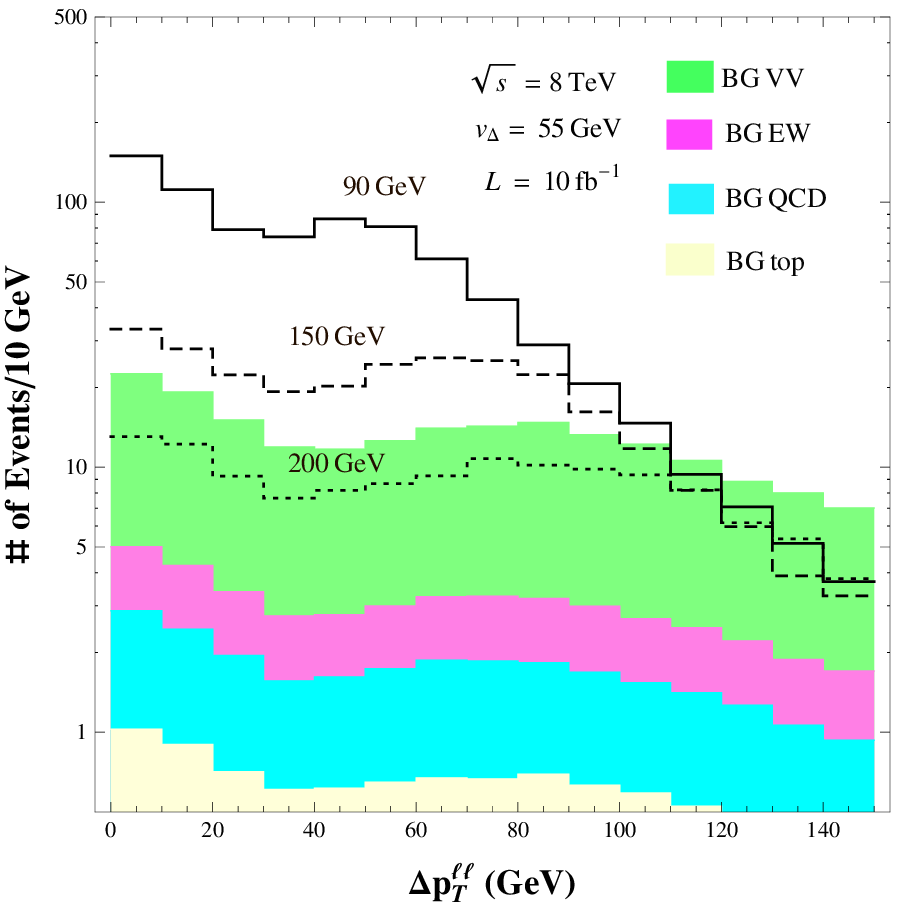}
\caption{Distributions of pre-selected events in $\Delta R^{\ell \ell}$ and $\Delta p_T^{\ell \ell}$, 
assuming $\sqrt{s}=8$ TeV, $v_{\Delta}=55$ GeV and $L = 10$ fb$^{-1}$.  
The doubly charged Higgs boson mass $m_{H^{\pm\pm}}=90, 150, 200$ GeV.   
The bin size for the $\Delta R^{\ell \ell}$ distribution is $0.1$.}
\label{FIG:lldistLow}
\end{figure}

We first examine the low-mass region where $m_{H^{\pm\pm}} \le 200$ GeV.
The left panel of FIG.~\ref{FIG:lldistLow} shows the distributions in $\Delta R^{\ell\ell}$ of 
the two leptons, defined by $\Delta R^{\ell\ell}=\sqrt{(\Delta \eta^{\ell\ell})^2+(\Delta \phi^{\ell\ell})^2}$, 
where $\Delta \phi^{\ell \ell}$ is the angle between the transverse momenta of the two leptons.  The right panel shows the distribution in the difference of the transverse momenta of the two leptons, 
$\Delta p_T^{\ell\ell}=|\vec{p}_T^{\, \ell_1}-\vec{p}_T^{\, \ell_2}|$. 
For the low-mass signals, the doubly charged Higgs boson should be energetic and 
thus the outgoing leptons tend to be in the same direction. 
Therefore, the signal events have smaller $\Delta R^{\ell\ell}$ and $\Delta p_T^{\ell\ell}$. 

After the pre-selection cut, we employ the following additional selection cuts:
\begin{subequations}
\label{Cut:NoJetLow}
\begin{align}
 & \Delta R^{\ell\ell}	<  2.4, 	  			\label{Cut:NoJetLow-1} \\
 & \Delta p_T^{\ell\ell} <  200~\text{GeV},			\label{Cut:NoJetLow-2} \\
 & M_{\ell\ell}          <  m_{H^{\pm\pm}}^{} + 10~\text{GeV}, 	\label{Cut:NoJetLow-3}
\end{align}
\end{subequations}
for the low-mass analysis. 
Since the $\Delta p_T^{\ell\ell}$ cut is correlated with the $\Delta R^{\ell\ell}$ cut, 
we here take a rather loose criterion for the $\Delta p_T^{\ell\ell}$ cut.

\begin{figure}[tb]
\centering
\includegraphics[scale=.85]{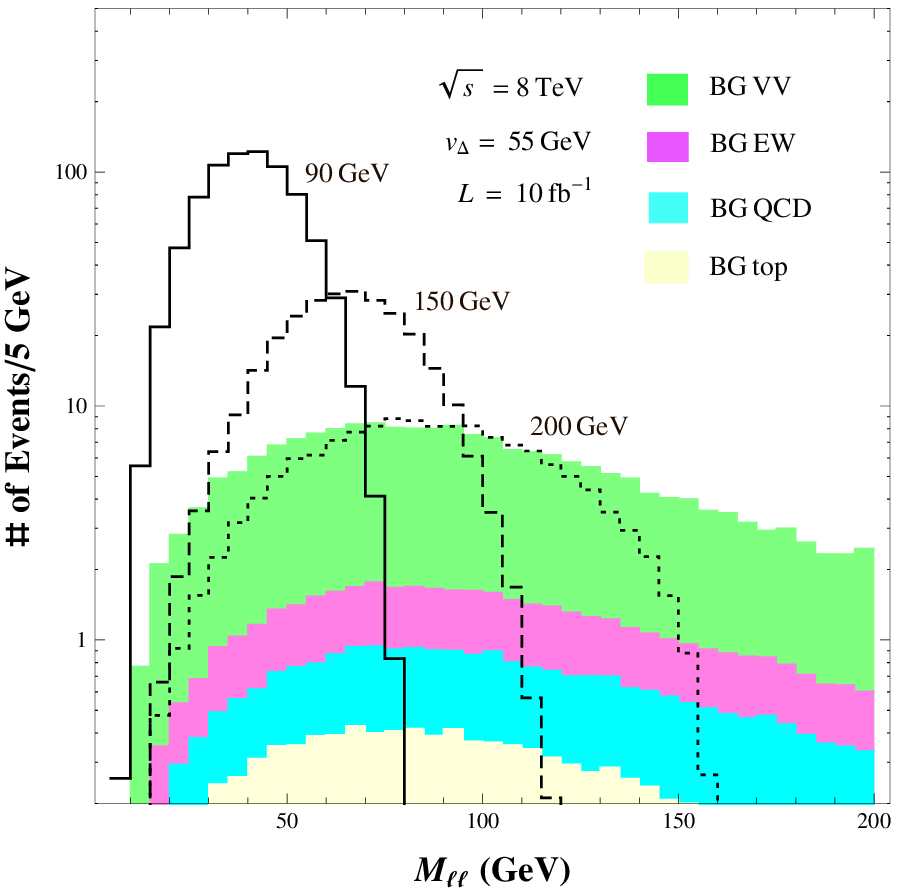}
\hspace{0.5cm}
\includegraphics[scale=.85]{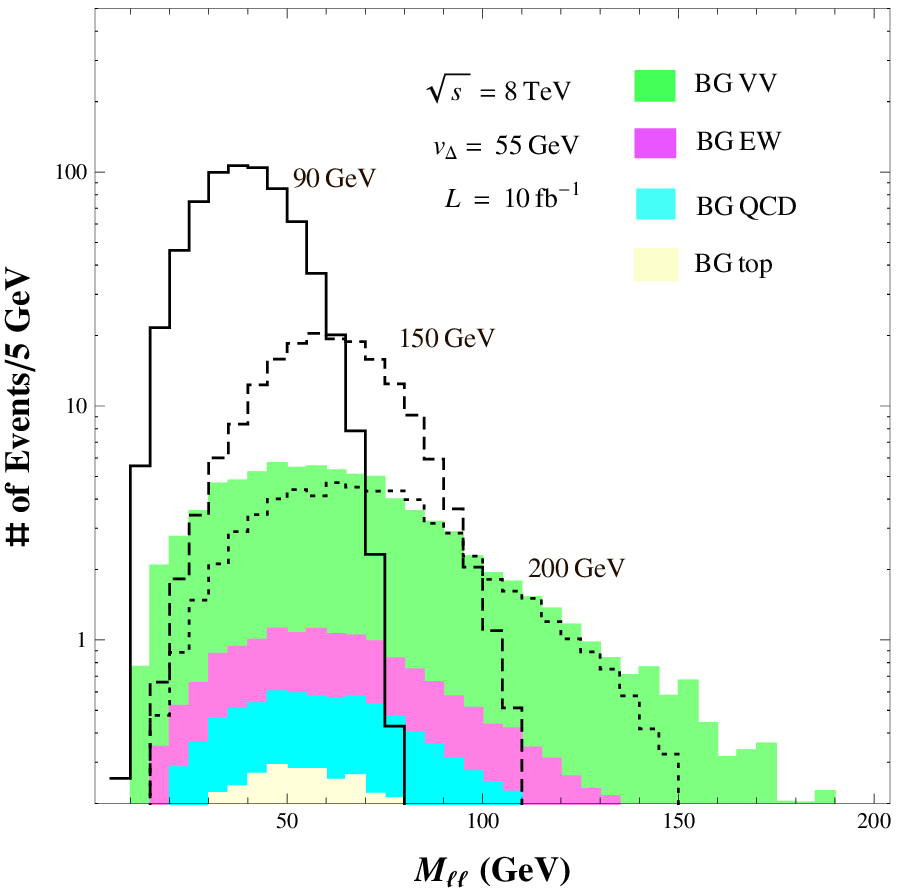}
\caption{Distributions of pre-selected events in the invariant mass $M_{\ell\ell}$ 
before (left) and after (right) the selection cuts of Eq.~\eqref{Cut:NoJetLow}, 
assuming $\sqrt{s}=8$ TeV, $v_{\Delta}=55$ GeV and $L = 10$ fb$^{-1}$. 
The doubly charged Higgs boson mass $m_{H^{\pm\pm}}=90, 150, 200$ GeV.}
\label{FIG:IMdistLow}
\end{figure}

The invariant mass distributions of pre-selected events before and after the selection cuts of Eq.~\eqref{Cut:NoJetLow} are given in the left and right panels of FIG.~\ref{FIG:IMdistLow}, respectively.  These selection cuts reduce the background events with a higher invariant mass. 
An invariant mass cut can further enhance the significance of signals in the case with 
a lighter doubly charged Higgs boson. 
In this case, one can extract the doubly charged Higgs boson mass by fitting to the entire distribution with sufficient statistics. 

\begin{table}[tb]
\centering
\begin{tabular}{|c||c||c|c|c|c|c||c|} \hline
w/o jet tagging
& $m_{H^{\pm\pm}}^{}\! =\! (90,150,200)$\! GeV 
       & DY & $VV$ & $t\bar{t}$ & EW & QCD & $\mathcal S$ 
\\ \hline \hline
pre-selection 
& (885,279,142)  & 5.8 & 180 & 12.0 & 25.1 & 19.6 & (37.4,15.7,8.5) \\ \hline 
$\Delta R^{\ell\ell} < 2.4$  
& (723,214,89.8) & 0 & 94.7 & 6.5 & 11.9 & 7.3 & (42.9,16.0,7.4) \\ \hline 
$\Delta p_T^{\ell\ell} < 200$~GeV  
& (715,212,87.4) & 0 & 90.6 & 6.1 & 11.0 & 6.9 & (43.3,16.1,7.4) \\ \hline
$M_{\ell \ell}\! <\! 210$\! \text{GeV} 
& (---,---,87.4) & 0 & 89.5 &  6.0 & 10.7 & 6.7 & (---,---,7.4) \\ \hline
$M_{\ell \ell}\! <\! 160$\! \text{GeV} 
& (---,212,---) & 0 & 86.5 &  5.7 & 10.2 & 6.4 & (---,16.4,---) \\ \hline
$M_{\ell \ell}\! <\! 100$\! \text{GeV} 
& (715,---,---) & 0 & 67.6 &  4.3 & 7.7 & 4.8 & (46.8,---,---) \\ \hline
\end{tabular}
\caption{Number of signal and background events left at each stage of cuts in Eq.~\eqref{Cut:NoJetLow}, 
assuming $\sqrt{s}=8$ TeV, $v_{\Delta}=55$ GeV and $L=10$ fb$^{-1}$.  
No forward jet tagging is imposed, but more than two jets are required in the pre-selection cut. \label{low}}
\label{Tab:NoJetLow}
\end{table}

The results of the signal/background reduction at each stage of the cuts are summarized in TABLE~\ref{Tab:NoJetLow}. 
We show the expected numbers of events and significance $\mathcal S$ for an integrated luminosity $L=10$~fb$^{-1}$ for each process.
Again, all the pair production, the VBF, and the associated 
production processes are taken as the signal events. 

For $m_{H^{\pm\pm}}^{} \lesssim 150$ GeV, the requirement of collinearity in 
the same-sign dileptons in Eqs.~\eqref{Cut:NoJetLow-1} and \eqref{Cut:NoJetLow-2} 
can reduce the number of background events while keeping most of the signal events, as seen in the increase of the significance. 
For $m_{H^{\pm\pm}}^{} \gtrsim 150$ GeV, the significance can hardly improve or even reduce. 
We will show later the required luminosity for a 5-sigma discovery as 
a function of the doubly charged Higgs boson mass based on this analysis.

\subsubsection{High-mass analysis}

\begin{figure}[tb]
\centering
\includegraphics[scale=.85]{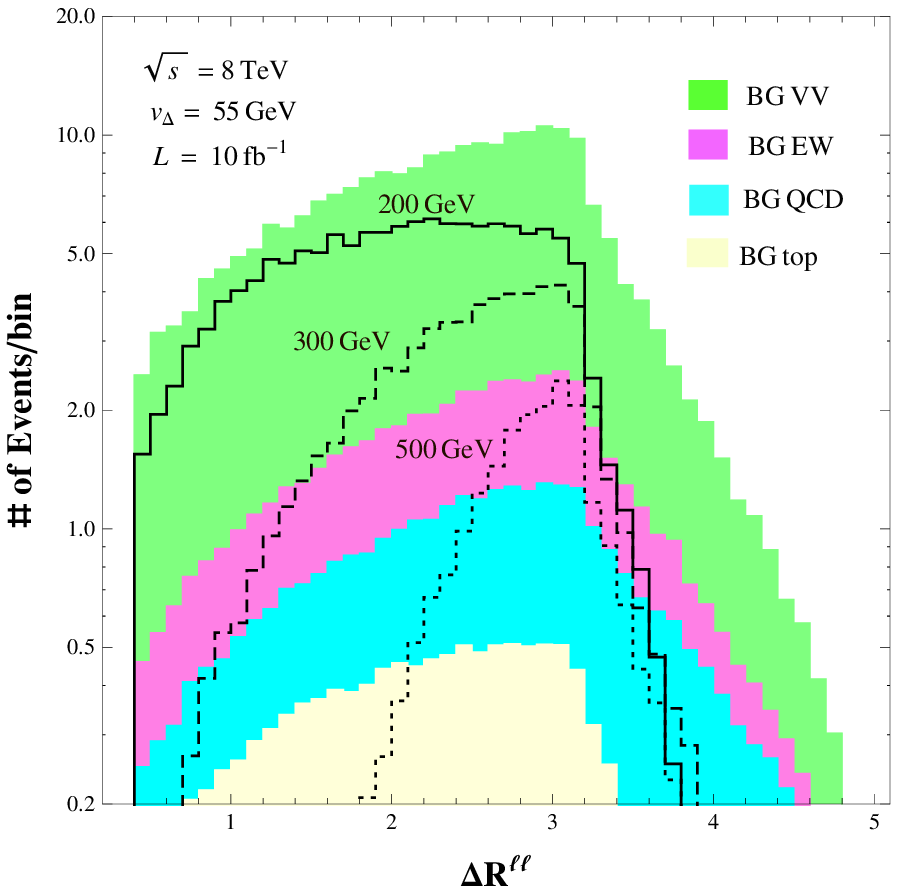}
\hspace{0.5cm}
\includegraphics[scale=.85]{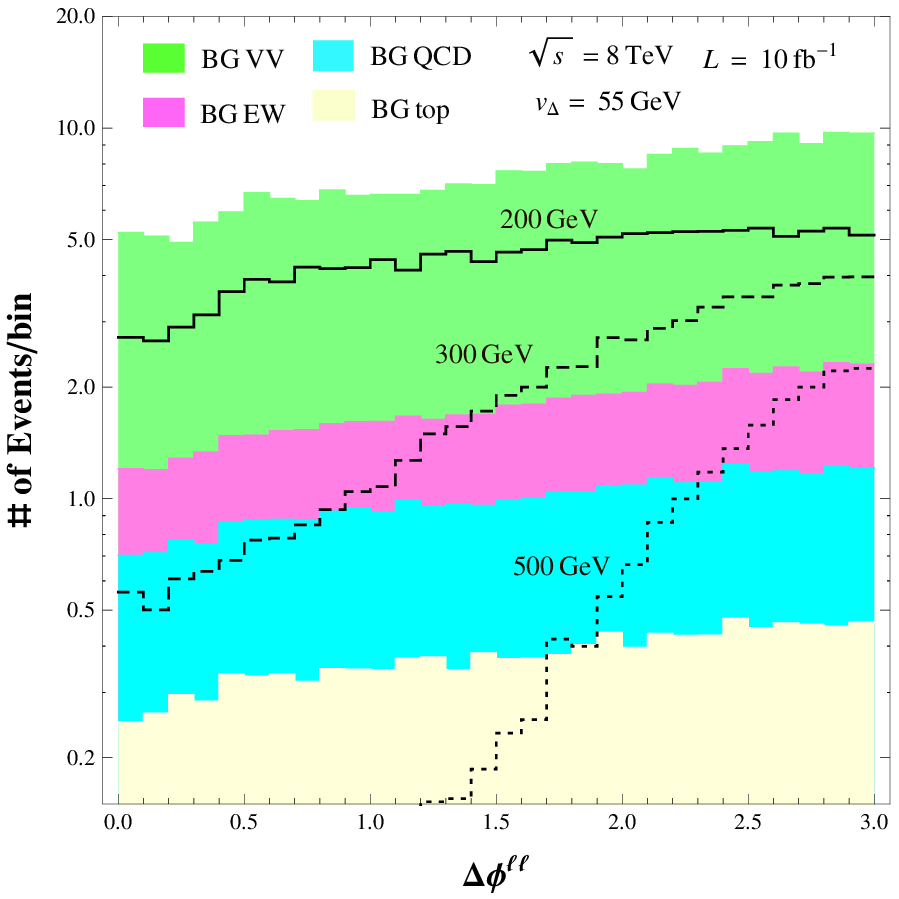}
\caption{Distributions of pre-selected events in $\Delta R^{\ell \ell}$ and $\Delta \phi^{\ell \ell}$, 
assuming $\sqrt{s}=8$ TeV, $v_{\Delta}=55$ GeV and $L = 10$ fb$^{-1}$.  
The doubly charged Higgs boson mass $m_{H^{\pm\pm}}=200, 300, 500$ GeV.  The bin size is $0.1$ for both distributions.}
\label{FIG:lldistHigh}
\end{figure}

We now consider the region where $m_{H^{\pm\pm}} \ge 200$ GeV.
In FIG.~\ref{FIG:lldistHigh}, we show the distribution for 
$\Delta R^{\ell\ell}$ of the two same-sign dileptons (left panel), 
and that for $\Delta \phi^{\ell\ell}$, the difference in the azimuthal 
angles (right panel). 
For the high-mass signals, the doubly charged Higgs boson from the VBF process would be produced nearly at rest.  The energetic leptons from the weak boson decays 
follow the parent (back-to-back) directions.
Therefore, the signal events have relatively large $\Delta \phi^{\ell\ell}$ 
(close to $\pi$). 

After the pre-selection cut, we apply the following additional selection cuts:
\begin{subequations}
\label{Cut:NoJetHigh}
\begin{align}
 & \Delta \phi^{\ell\ell}	>  \pi/2, 	  		\label{Cut:NoJetHigh-1} \\
 & \Delta R^{\ell\ell}		<  3.5,				\label{Cut:NoJetHigh-2} \\
 & M_{\ell\ell}         	>  m_{H^{\pm\pm}}^{}/4, 	\label{Cut:NoJetHigh-3}
\end{align}
\end{subequations}
for the high-mass analysis. 

\begin{figure}[tb]
\centering
\includegraphics[scale=.85]{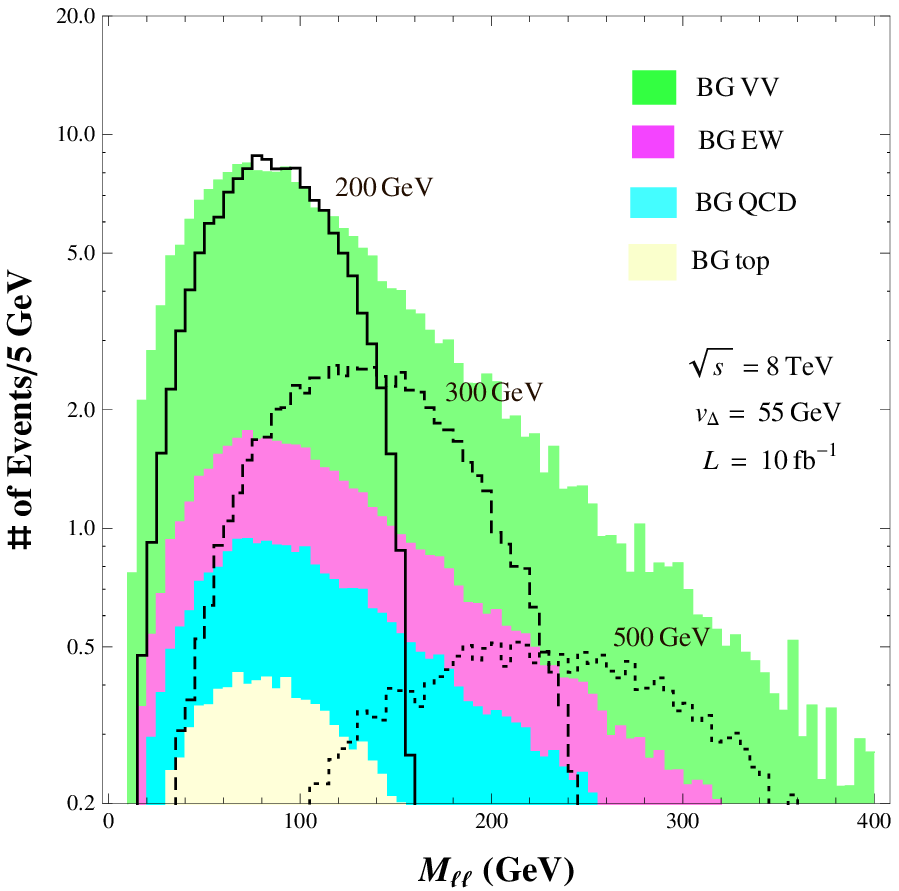}
\hspace{0.5cm}
\includegraphics[scale=.85]{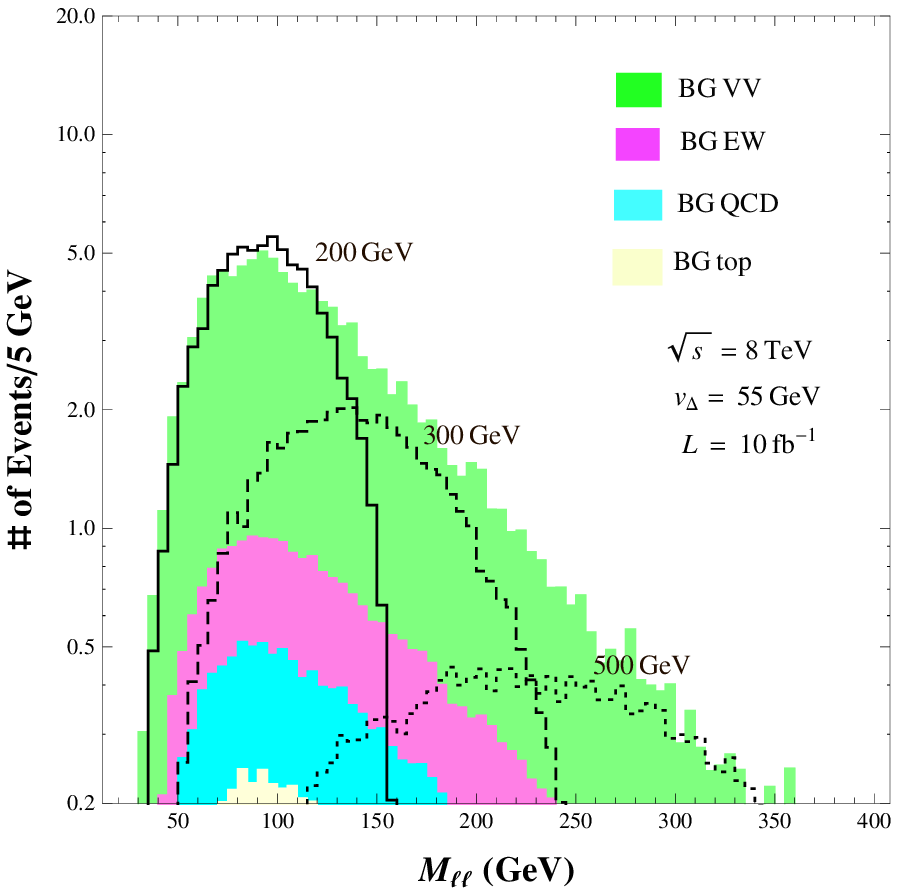}
\caption{Distributions of pre-selected events in the invariant mass $M_{\ell\ell}$ 
before (left) and after (right) the selection cuts in Eq.~\eqref{Cut:NoJetHigh}, 
assuming $\sqrt{s}=8$ TeV, $v_{\Delta}=55$ GeV and $L = 10$ fb$^{-1}$. 
The doubly charged Higgs boson mass $m_{H^{\pm\pm}}=200, 300, 500$ GeV.}
\label{FIG:IMdistHigh}
\end{figure}

The invariant mass distributions of the pre-selected events with the same-sign dileptons before and after the selection cuts in Eq.~\eqref{Cut:NoJetHigh} are given in the left and right panels of FIG.~\ref{FIG:IMdistHigh}, respectively. 
Apparently, in the high-mass case, the invariant mass cut cannot enhance the signal significance much.  Also, it may be difficult to determine the doubly charged Higgs boson mass from the invariant mass distribution because of the low statistics.

%
\begin{table}[tb]
\centering
\begin{tabular}{|c||c||c|c|c|c|c||c|} \hline
w/o jet tagging
& $m_{H^{\pm\pm}}^{}\! =\! (200,300,500)$\! GeV 
       & DY & $VV$ & $t\bar{t}$ & EW & QCD & $\mathcal S$ 
\\ \hline \hline
pre-selection 
& (142.,65.7,21.5) & 5.8 & 180. & 12.0 & 25.1 & 19.6 & (8.5,4.0,1.4) \\ \hline 
$\Delta \phi^{\ell\ell} >  \pi/2$  
& (81.2,49.8,20.1) & 2.9 & 105. & 6.7 & 15.1 & 11.0 & (6.3,4.0,1.7) \\ \hline  
$\Delta R^{\ell\ell} <  3.5$  
& (79.3,47.7,18.4) & 2.9 & 93.1 & 6.1 & 13.0 & 7.7 & (6.5,4.1,2.0) \\ \hline 
$M_{\ell \ell}\! >\! 50$\! \text{GeV} 
& (76.2,---,---)  & 0. & 89.8 &  6.0 & 12.7 &  7.4 & (6.5,---,---) \\ \hline
$M_{\ell \ell}\! >\! 75$\! \text{GeV} 
& (---,44.7,---) & 0. & 80.7 &  5.4 & 11.7 &  6.8 & (---,4.2,---) \\ \hline
$M_{\ell \ell}\! >\! 125$\! \text{GeV} 
& (---,---,16.8) & 0. & 37.9 &  2.9 & 6.8 & 4.0 & (---,---,2.2) \\ \hline
\end{tabular}
\caption{Number of signal and background events left at each stage of cuts in Eq.~\eqref{Cut:NoJetHigh}, assuming $v_{\Delta}=55$ GeV and $L=10$ fb$^{-1}$ for the LHC running at 8 TeV.  No forward jet tagging is imposed, but more than two jets are required in the pre-selection cut. \label{high}}
\label{Tab:NoJetHigh}
\end{table}

The results of the signal/background reduction at each stage of the cuts are summarized in TABLE~\ref{Tab:NoJetHigh}, assuming an integrated luminosity of $10$~fb$^{-1}$.
We again note that all the pair production, the VBF, and the associated 
production processes are taken into account as the signal events. 

For $m_{H^{\pm\pm}}^{}\gtrsim 300$ GeV, the back-to-back cut in Eq.~\eqref{Cut:NoJetHigh-1} rejects more background than signal.  
When the mass of the doubly charged Higgs boson increaces, 
the significance can be better than the low mass analysis. 
We will also give the required luminosity for a 5-sigma discovery in this case later. 

\subsection{Event analysis with forward jet tagging}

We next focus on the doubly charged Higgs boson signal from 
the VBF process, which becomes the dominant production mechanism 
when the triplet VEV is sufficiently large. 
The signal events consist of same-sign dileptons with two high-$p_T^{}$ jets in the forward region. 
As shown in FIG.~\ref{FIG:Xsec}, the VBF process 
becomes most important for a heavier doubly charged Higgs boson. 
Therefore, we consider $v_\Delta^{}=55$ GeV and 
$m_{H^{\pm\pm}} = 150$, $300$ and $500$ GeV as examples in this subsection. 
The LHC running at $8$ TeV is also assumed in this subsection. 

We again demand the same-sign dileptons and two or more jets with 
no other high $p_T^{}$ leptons as the pre-selection cut. 
Therefore, the distributions for various kinematical variables of the 
leptons are the same as those presented in the previous subsection. 

\begin{figure}[tb]
\centering
\includegraphics[scale=0.85]{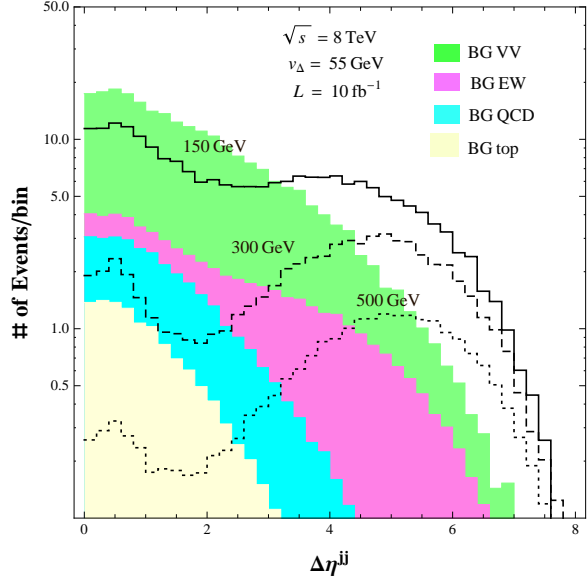}
\caption{Distribution of pre-selected events in $\Delta \eta^{jj}$, 
assuming $\sqrt{s}=8$ TeV, $v_{\Delta}=55$ GeV and $L = 10$ fb$^{-1}$. 
The bin size is $0.2$ for the distribution.}
\label{Deletajj}
\end{figure}

The most characteristic feature of the VBF is that 
the signal event contains two high-$p_T^{}$ jets in the forward region.
In FIG.~\ref{Deletajj}, we show the distributions of the signal and background events in the difference of the pseudorapidities of the two highest $p_T^{}$ jets. 
Clearly, the signal events tend to have large $\Delta \eta^{jj}$. 
Therefore, the forward jet tagging cut in the pseudorapidity difference 
can reduce the background events significantly. 

After the pre-selection cut, we require the following additional selection cuts:
\begin{subequations}
\label{Cut:TagJet}
\begin{align}
 & \Delta \eta^{jj}		>  3.5, 	  		\label{Cut:TagJet-1} \\
 & \Delta \phi^{\ell\ell}	>  \pi/2, 	  		\label{Cut:TagJet-2} \\
 & \Delta R^{\ell\ell}		<  3.5,				\label{Cut:TagJet-3} \\
 & M_c(WW)         		>  m_{H^{\pm\pm}}^{}/2, 	\label{Cut:TagJet-4}
\end{align}
\end{subequations}
where the cluster transverse mass $M_c(WW)$ is defined in Eq.~\eqref{Eq:MC}.  

\begin{figure}[tb]
\centering
\includegraphics[scale=.85]{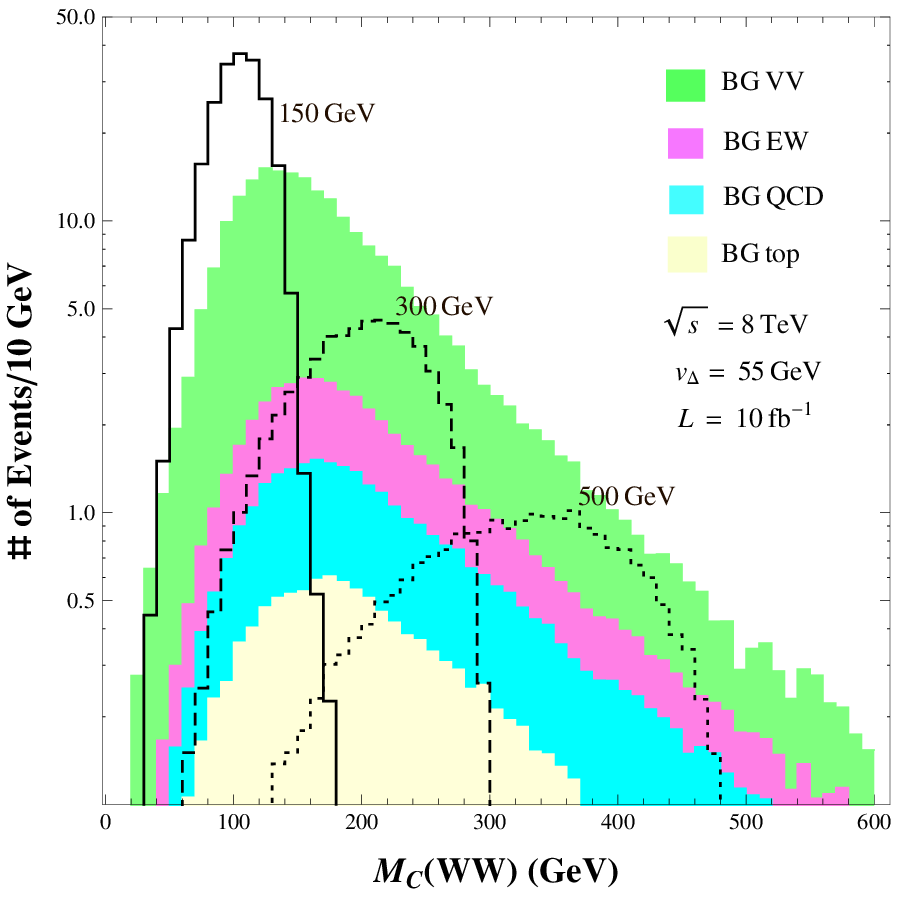}
\hspace{0.5cm}
\includegraphics[scale=.85]{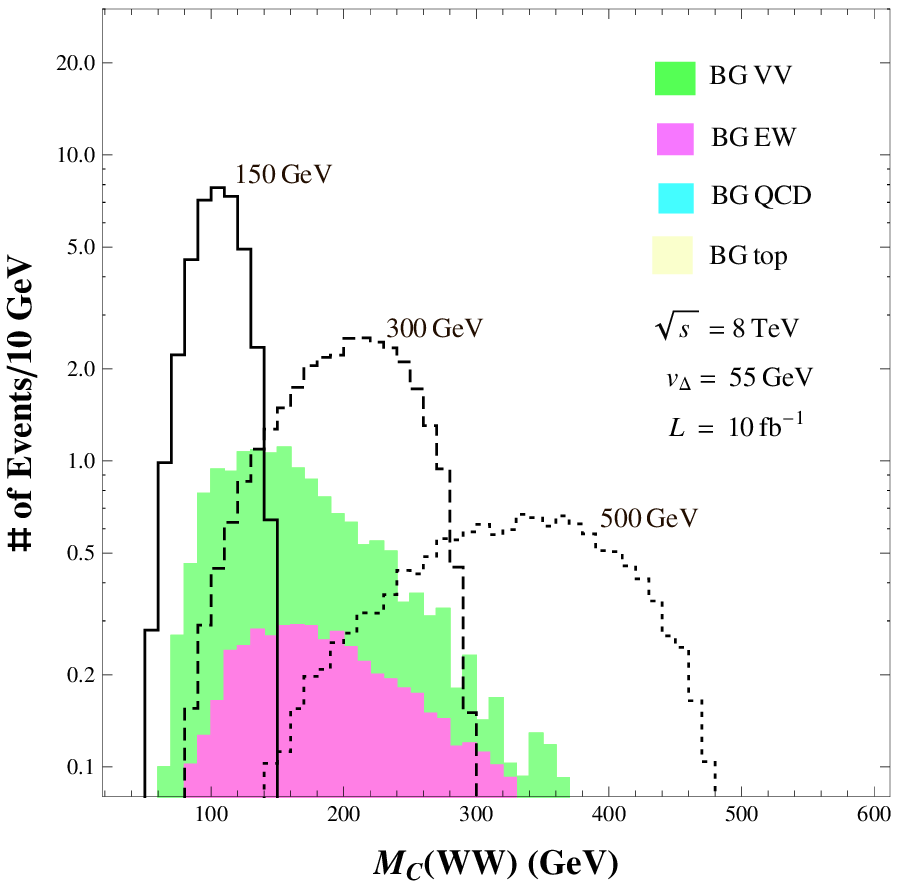}
\caption{Distributions of preselected events in the cluster transverse mass $M_c (WW)$ 
before (left) and after (right) imposing the cuts in Eq.~\eqref{Cut:TagJet}, 
assuming $\sqrt{s}=8$ TeV, $v_{\Delta}=55$ GeV and $L = 10$ fb$^{-1}$. 
The doubly charged Higgs boson mass is taken to be $m_{H^{\pm\pm}}=150, 300, 500$ GeV. 
}
\label{FIG:MC}
\end{figure}

In FIG.~\ref{FIG:MC}, we show the distributions of the pre-selected events in the cluster transverse mass. 
The left and right panels respectively show the results before and after the first three selection cuts in Eq.~\eqref{Cut:TagJet} are imposed. 
After the selection cuts, the background events tend to have smaller $M_c(WW)$. 
Therefore, the cut in the cluster transverse mass in Eq.~\eqref{Cut:TagJet-4} 
further improves the significance.

\begin{table}[tb]
\centering
\begin{tabular}{|c||c||c|c|c|c|c||c|} \hline
w/ jet tagging
& $m_{H^{\pm\pm}}^{}\! =\! (150,300,500)$\! GeV 
       & DY & $VV$ & $t\bar{t}$ & EW & QCD & $\mathcal S$ 
\\ \hline \hline
pre-selection 
& (279,65.7,21.5)  & 5.8 & 180 & 12.0 & 25.1 & 19.6 & (15.7,4.1,1.4) \\ \hline
$\Delta \eta^{jj} >  3.5$  
& (74.9.,40.5,16.5)  & 0 & 19.9 & 0.2 & 9.7 & 1.0 & (10.0,6.2,2.8) \\ \hline  
$\Delta \phi^{\ell\ell} >  \pi/2$  
& (38.7,32.5,15.5) & 0 & 12.5 & 0.1 & 6.0 & 0.6 & (7.1,6.1,3.2) \\ \hline 
$\Delta R^{\ell\ell} <  3.5$  
& (38.3,31.1,14.2) & 0 & 11.0 & 0.1 & 5.2 & 0.3 & (7.4,6.2,3.1) \\ \hline 
$M_{c}(WW)\! >\! 75$\! \text{GeV} 
& (36.0,---,---.)  & 0 & 10.3 &  0.1 & 5.2 &  0.3 & (7.1,---,---) \\ \hline
$M_{c}(WW)\! >\! 150$\! \text{GeV} 
& (---,26.2,---)   & 0 & 5.6 &  0.1 & 3.8 &  0.2 & (---,6.5,---) \\ \hline
$M_{c}(WW)\! >\! 250$\! \text{GeV} 
& (---,---,11.3)   & 0 & 1.2 &  0 & 1.6 & 0.1 & (---,---,4.7) \\ \hline
\end{tabular}
\caption{Number of signal and background events left at each stage of cuts, assuming $v_{\Delta}=55$ GeV and $L=10$ fb$^{-1}$ for the LHC running at 8 TeV.  More than two jets are required at the pre-selection. \label{jtag}}
\label{Tab:JetTag}
\end{table}

The results of the signal/background analysis are given in TABLE~\ref{Tab:JetTag}. 
The expected numbers of events are scaled for the integrated luminosity of
$10$~fb$^{-1}$ for each process.
All the pair production, the VBF, and the associated 
production processes are treated as the signal events. 
For a relatively heavy doubly charged Higgs boson, the forward jet tagging cut 
substantially improves the significance. 
Although the significance is reduced by the selection cut for a light doubly 
charged Higgs boson, such a parameter region can be well tested by the analysis 
presented in the previous subsection.
A remark on how the significance scales with the triplet VEV is in order. 
If the signal comes purely from the VBF process, the cross section 
is proportional to $v_\Delta^2$.  Therefore, the expected significance using forward jet tagging varies linearly with $v_\Delta$.

\subsection{Discovery reach}

\begin{figure}[t]
\includegraphics[scale=0.9]{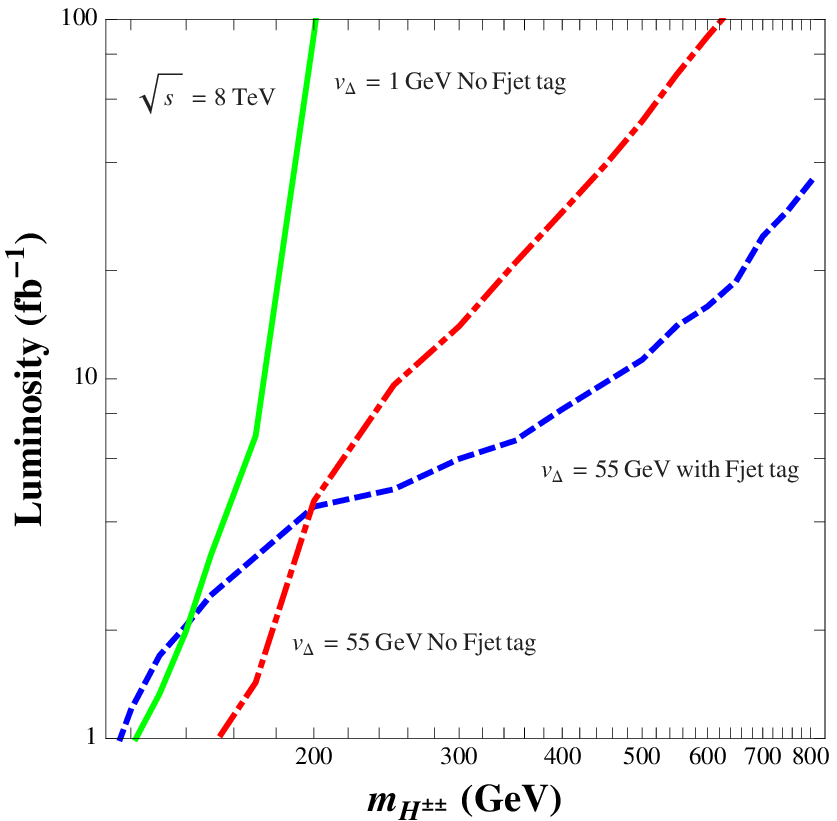}
\includegraphics[scale=0.9]{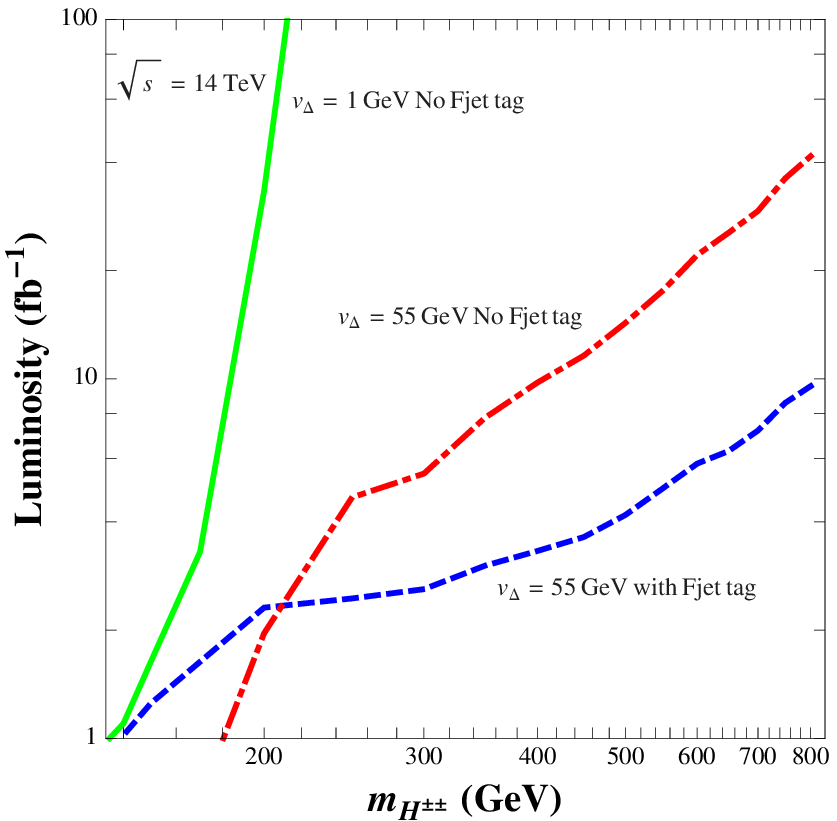}
\caption{The required luminosity for a 5-sigma discovery of $H^{\pm\pm}$ at the LHC as a function of $m_{H^{\pm\pm}}$, where the kinematical cuts explained in the previous section are imposed and ``Fjet tag" means forward jet tagging.  The left (right) plot is for the collision energy of 8 TeV (14 TeV).
\label{luminosity_5sigma}
}
\end{figure}

We now present in FIG.~\ref{luminosity_5sigma} the required luminosity for
a 5-sigma discovery of the doubly charged Higgs boson in the $W^\pm W^\pm$ decay channel with the same-sign dilepton signature in the final state as a function of $m_{H^{\pm\pm}}$.
The left (right) plot shows the required luminosities for two different values of $v_\Delta$ in the case of 8 TeV (14 TeV) collisions. 
The kinematical distributions of leptons and jets for the $14$ TeV 
case have almost the same shape as the $8$ TeV case, differing only in the overall magnitude.  Thus, one can apply the same analysis as for the $8$ TeV case shown in Sec.~\ref{sec:simulation}.

In FIG.~\ref{luminosity_5sigma}, the red dot-dashed curves indicate the required luminosity for the analysis without forward jet tagging and $v_{\Delta}=55$GeV.  
In order to optimize the significance, the kinematical cuts in Eqs.~\eqref{Cut:NoJetLow-1}, \eqref{Cut:NoJetLow-2} and \eqref{Cut:NoJetLow-3} are applied in the low-mass region $m_{H^{\pm\pm}} \leq 200$GeV, while those in Eqs.~\eqref{Cut:NoJetHigh-1}, \eqref{Cut:NoJetHigh-2} and \eqref{Cut:NoJetHigh-3} are applied in the high-mass region $m_{H^{\pm\pm}} > 200$GeV.
This explains the kinks in the curves at $m_{H^{\pm\pm}}=200$ GeV.

The green solid curves are for the analysis without forward jet tagging and $v_{\Delta}=1$ GeV, assuming the cross sections are scaled as $v_\Delta^{2}$ for the VBF and the $H^{\pm \pm}W^{\mp}$ production processes.
The case of $v_{\Delta}=1$ GeV corresponds to the limit where the VBF and  
the associated production processes become negligible and the pair production processes take over.
The required luminosity rapidly increases for $m_{H^{\pm\pm}} \gtrsim 180$GeV as the pair production cross sections become too small.

The blue dashed curves give the required luminosity for the analysis with forward jet tagging and $v_{\Delta}=55$ GeV.
We apply the kinematical cuts in Eqs.~\eqref{Cut:TagJet-1}, \eqref{Cut:TagJet-2}, \eqref{Cut:TagJet-3} and \eqref{Cut:TagJet-4} for $m_{H^{\pm\pm}} > 200$ GeV but only Eq.~(\ref{Cut:TagJet-1}) for $m_{H^{\pm\pm}} \leq 200$ GeV because the
other cuts reduce the signal events in this mass region.
%
%
If we take $v_{\Delta}=1$ GeV, the analysis with forward jet tagging does not work since the cross section of VBF becomes too small and the luminosity will be out of LHC scope for the entire mass region.

The figures show that the analysis with forward jet tagging is more effective 
for $v_{\Delta}=55$ GeV and $m_{H^{\pm\pm}} \gtrsim 200$ GeV where the VBF becomes dominant.
On the other hand, the analysis without forward jet tagging is effective for $m_{H^{\pm\pm}}<200$ GeV since the pair production processes are dominant in this region.
We thus conclude that the discovery potential for $H^{\pm \pm}$ at the LHC is promising through the $W^\pm W^\pm$ mode with the forward jet tagging analysis when $v_{\Delta}$ is sufficiently large, such as $55$ GeV as allowed in the Georgi-Machacek model.
The doubly charged Higgs boson with a mass of 450 GeV (800 GeV) can be discovered at the LHC with an integrated luminosity of $10$ fb${}^{-1}$ at 8 TeV (14 TeV).
We also find the discovery potential for a light $H^{\pm \pm}$ is in the scope of LHC even if $v_\Delta \lesssim 1$ GeV where only the pair production processes dominate.
The mass of 180 GeV (190 GeV) can be discovered for an integrated luminosity of $10$ fb${}^{-1}$ at 8 TeV (14 TeV).

%
%

\section{Conclusions and discussions
\label{sec:summary}}

In this paper, we have studied the doubly charged Higgs boson production at the Large Hadron Collider in the Higgs triplet model (HTM) and the Georgi-Machacek model.
We focus on the scenario in which the decay channel of a pair of weak gauge bosons is dominant for the doubly changed Higgs boson due to a large vacuum expectation value (VEV) of the Higgs triplet field.

In this scenario, the doubly charged Higgs boson is produced via the pair production processes $pp \rightarrow H^{++}H^{--}$ and $H^{\pm \pm} H^{\mp}$, the VBF process $pp \rightarrow H^{\pm \pm}jj$, and the associated production process $pp \rightarrow H^{\pm \pm} W^{\mp}$.  We take two benchmark values for the triplet VEV, $v_\Delta$: 1 GeV and 55 GeV.  The former corresponds to the limit when the pair production processes are dominant, whereas the latter is the upper limit of the triplet VEV for $m_{H^{\pm\pm}} \sim {\cal O}(100 \mbox{ GeV})$ in the Georgi-Machacek model.
The pair production rates are independent of the triplet VEV, $v_\Delta$.
On the other hand, the cross sections of the VBF and the associated production are proportional to $v_\Delta^2$.
The VBF production becomes dominant for a heavy doubly charged Higgs boson ($m_{H^{\pm\pm}} \gtrsim 100$ GeV) when $v_\Delta$ is sufficiently large. 

Taking $v_\Delta=55$ GeV, we present in FIG.~\ref{FIG:Xsec} the cross sections of various doubly charged Higgs boson production processes for the LHC operating at 8 TeV and 14 TeV.
We have performed a detailed simulation of the signals and the backgrounds at the LHC.  Only the distributions for the 8 TeV collisions are presented because the 14 TeV case has virtually the same distribution shapes, differing only in magnitude.
In our scenario, the doubly charged Higgs boson decays into a pair of same-sign $W$ bosons, each of which further decays into lepton pairs, resulting in a pair of same-sign dileptons as the signature of lepton number violation.  The $W$ bosons can be off-shell, depending on the doubly charged Higgs boson mass.
%
%
We have carried out the analysis without forward jet tagging in both high-mass and low-mass regions separately and discussed the corresponding cuts. 
We have also performed the analysis with forward jet tagging to improve significance in the high-mass region where the VBF production mechanism becomes dominant.
The significance in these analyses is summarized in TABLEs~\ref{Tab:NoJetLow}, \ref{Tab:NoJetHigh} and \ref{Tab:JetTag}, assuming an integrated luminosity of $10$ fb$^{-1}$. 
We find that the analysis with forward jet tagging is most effective 
for $v_{\Delta}=55$ GeV and $m_{H^{\pm\pm}} \gtrsim 200$ GeV.
On the other hand, the analysis without jet tagging is effective for $m_{H^{\pm\pm}}<200$ GeV since the cross sections of pair production processes are large in this region.

Finally, we have shown the required luminosity for discovering the doubly charged Higgs boson at the 5-sigma level in FIG.~\ref{luminosity_5sigma}. 
The discovery potential for heavy $H^{\pm \pm}$ at the LHC is significantly large for $v_{\Delta}=55$ with the forward jet tagging analysis.
With an integrated luminosity $10$ fb${}^{-1}$, the LHC operating at 8 TeV (14 TeV) can reach up to the mass of 450 GeV (800 GeV).
The discovery potential for light $H^{\pm \pm}$ is also in the scope of LHC even if $v_\Delta \lesssim 1$ GeV where the only the pair production mechanisms contribute.
Also assuming $L=10$ fb${}^{-1}$, the LHC running at 8 TeV (14 TeV) can reach up to the mass of 180 GeV (190 GeV).

\acknowledgments
We thank C.-M. Kuo and H. Yokoya for some useful suggestions regarding experimental capacities.
We also thank H. Sugiyama for valuable comments.
This research was supported in part by the National Science Council of Taiwan, R.~O.~C. under Grant Nos.~NSC-100-2628-M-008-003-MY4 and NSC-100-2811-M-002-090.



\end{document}